\pdfoutput=1
\documentclass[a4paper]{article}
\usepackage[a4paper,margin=3.00cm]{geometry}

\usepackage{graphicx}

\usepackage{booktabs}
\usepackage{lmodern}
\usepackage[T1]{fontenc}

\usepackage{amsmath}
\usepackage{amssymb} 
\usepackage{cite}

\usepackage[hidelinks]{hyperref}

\hypersetup{
    pdfauthor={Nathan Clisby},
    pdftitle={High resolution Monte Carlo study of the Domb-Joyce model}
    }

\providecommand{\norm}[1]{\vert #1 \vert}
\providecommand{\ee}[1]{\ensuremath{\times 10^{#1}}}

\newcommand{\Z}{{\mathbb Z}}
\newcommand{\rimp}{R^2_{\text{imp}}}
\newcommand{\resq}{R_{\mathrm{E}}^2}

\newcommand{\rgsq}{R_{\mathrm{G}}^2}

\newcommand{\dde}{D_{\mathrm{E}}}
\newcommand{\ddg}{D_{\mathrm{G}}}
\newcommand{\aae}{a_{\mathrm{E}}}
\newcommand{\aag}{a_{\mathrm{G}}}

\newcommand{\avresq}{\langle R_{\mathrm{E}}^2 \rangle}
\newcommand{\avrgsq}{\langle R_{\mathrm{G}}^2 \rangle}
\newcommand\tstrut{\rule{0pt}{2.4ex}}
\newcommand\bstrut{\rule[-1.0ex]{0pt}{0pt}}

\usepackage{float}

\begin{document}

\title{High resolution Monte Carlo study of the Domb-Joyce model}
	
\author{Nathan Clisby \\
School of Mathematics and Statistics, \\
The University of Melbourne, Victoria 3010, Australia. \\
nclisby@unimelb.edu.au}

\date{May 3, 2017}

\maketitle

\begin{abstract}
We study the Domb-Joyce model of weakly self-avoiding walks on the
simple cubic lattice via Monte Carlo simulations. We determine to
    excellent
    accuracy the value
for the interaction parameter which results in an improved model for
    which the leading correction-to-scaling term has zero amplitude.
\end{abstract}

\section{Introduction}
\label{sec:intro}

The Domb-Joyce model of weakly self-avoiding walks has a long
history~\cite{Domb1973Selfavoidingwalks}. 
It consists of the set of random walks on the simple cubic lattice
$\Z^3$, with walks starting at the origin and taking steps to adjacent
neighbors, where there is a contact interaction which introduces an
energy penalty of
$w$ for each pair of visits to the same lattice site.
A walk of $N$ steps may be defined as a mapping $\omega$ from the
integers $0,1,\cdots,N$ to sites on $\Z^3$, with $\norm{\omega(i+1) -
\omega(i)} = 1$ $\forall i\in[0,N-1]$, and $\omega(i) \neq \omega(j)$
$\forall i \neq j$, and 
the energy due to overlaps of a walk $\omega$ is then given by
\begin{align}
    E(\omega) &= w \sum_{i<j} \delta_{\omega(i),\omega(j)}.
\end{align}
Note that if a walk visits a particular site exactly $k$
times, then the associated weight for that site will be equivalent to
$\binom{k}{2}$ pairwise interactions.
We then define the partition function as
\begin{align}
    Z_N &= \sum_{|\omega|=N} e^{-E(\omega)},
\end{align}
and the expectation of an observable $A$ is computed over the set of all simple random walks of
length $N$ as follows:
\begin{align}
    \langle A \rangle_N &= \frac{1}{Z_N} \sum_{|\omega|=N} A(\omega)
    e^{-E(\omega)}.
\end{align}
The Domb-Joyce model
interpolates between simple random
walks when $w=0$, and self-avoiding walks when $w=+\infty$.

Here we perform Monte Carlo simulations of Domb-Joyce walks via the
pivot
algorithm~\cite{Lal1969MonteCarlocomputer,Madras1988PivotAlgorithmHighly}
using a recent
implementation~\cite{Clisby2010Efficientimplementationpivot,Clisby2010AccurateEstimateCritical}
(which improved on earlier work by
Kennedy~\cite{Kennedy2002fasterimplementationpivot}).

We calculated the expected value of the two most common measures of
size, the squared end-to-end distance, $\resq$, and the squared radius
of gyration, $\rgsq$, which are defined as:
\begin{align}
    \label{eq:defre2}
    \resq & = \norm{\omega(N) - \omega(0)}^2;
\\
\label{eq:defrg2}
    \rgsq & = \frac{1}{2 (N+1)^2} \sum_{i,j}
    \norm{\omega(i)-\omega(j)}^2.
\end{align}

For any positive $w$ the Domb-Joyce model is in the same universality
class as self-avoiding walks, and the asymptotic behavior of these
observables for large $N$ is given by
\begin{align}
\avresq & = 
    \dde(w) N^{2 \nu} \left(1 +  \frac{\aae(w)}{N^{\Delta_1}} + \cdots \right) , \\
\avrgsq & = 
    \ddg(w) N^{2 \nu} \left(1 +  \frac{\aag(w)}{N^{\Delta_1}} + \cdots
    \right).
\end{align}
The critical exponent $\nu$, known as the Flory exponent, is a universal
quantity with value $\nu = 0.58759700(40)$~\cite{Clisby2016HydrodynamicRadiusForSAWs}. The
leading correction-to-scaling exponent $\Delta_1$ is also universal, and
the best estimate for it is $\Delta_1 = 0.528(8)$. The amplitudes
$\dde(w)$, $\ddg(w)$, $\aae(w)$, and $\aag(w)$ are non-universal
quantities which depend on the details of the model, including the value
of the weight $w$ and the lattice type. But, crucially for the present
study, the amplitude ratios $\dde(w)/\ddg(w) =
6.253531(10)$~\cite{Clisby2016HydrodynamicRadiusForSAWs} and $\aae(w)/\aae(w)$ are universal
quantities which are independent of $w$.

A key reason for the continuing interest in the Domb-Joyce model is the
observation that the addition of the interaction
parameter $w$ allows for the possibility of tuning it to a value such 
that the leading correction-to-scaling terms have negligible amplitude.
As $\aae(w)/\aae(w)$ is universal, if there is a value $w = w^*$ such
that $\aae(w^*) = 0$, then $\aag(w^*) = 0$ also. Thus, if $w^*$ can be
found to good accuracy it will ensure that the leading
correction-to-scaling term for each observable will have small
amplitude, thus enhancing convergence in the large-$N$ limit.
The Domb-Joyce model with $w = w^*$ is called an ``improved'' model due
to this feature.

Previously, Belohorec~\cite{Belohorec1997Renormalizationgroupcalculation} estimated that
$w^* = 0.506$ for the Domb-Joyce model, and 
Caracciolo et al.~\cite{Caracciolo2006Virialcoefficientsand} found that $w^* =
0.48(2)$ when studying virial coefficients for the self-avoiding walk
universality class.
More recently, Adamo and
Pelissetto~\cite{Adamo2016ImprovedModelPolymersHardSphereMixture} made a
detailed study of the Domb-Joyce model in the presence of hard spheres, and
obtained an estimate of $w^* = 0.486 \pm 0.003 \pm 0.005$, where the first error
is statistical, and the second comes from varying parameters which were used to
bias their fits.

Here we seek to build on earlier work and obtain an accurate estimate of $w^*$
to facilitate future Monte Carlo studies of properties of the self-avoiding walk
universality class.

\section{Monte Carlo simulation}
\label{sec:mc}

We sampled self-avoiding walks on $\Z^3$ over a range of lengths from one
thousand to ten million steps, for weights $e^{-w} =$ 0.50, 0.54, 0.57, 0.59,
0.60, 0.61, 0.63, 0.66, and 0.70 ($w =$ 0.6931, 0.6162, 0.5621, 0.5276, 0.5108,
0.4943, 0.4620, 0.4155, and  0.3567 respectively).  We used a variant of the
SAW-tree implementation of the pivot algorithm as described
in~\cite{Clisby2010Efficientimplementationpivot} to collect data for the
observables $\avresq$ and $\avrgsq$.

The SAW-tree implementation was adapted to count the number of intersections
created when a pivot move is performed, whereas for self-avoiding walks one only
needs to know if an intersection occurs. To initialize the system we used a
variant of the pseudo\_dimerize procedure, and to eliminate any initialization
bias we then performed approximately $20 N$ successful pivots before collecting
any data.

We sampled pivot sites uniformly at random along the chain, and with
the pivot symmetry operations sampled uniformly at random from amongst the 47 lattice
symmetries of $\Z^3$ that do not correspond to the identity.

After initialization, we started collecting data for
our observables for each time step, and aggregated the results in
batches of $10^8$.

The computer experiment was run for 130 thousand CPU hours on Dell PowerEdge
FC630 machines with Intel Xeon E5-2680 CPUs. (Run in hyperthreaded mode, so in
fact 260 thousand thread hours were used.) In total there were $5.7 \times 10^5$
batches of $10^8$ attempted pivots, and thus there were a grand total of $5.7
\times 10^{13}$ attempted pivots across all walk sizes and values of the
parameter $w$.

Our data for $\avresq$,$\avrgsq$, and $\avresq/\avrgsq$ for different values of
$w$ are collected in Tables~\ref{tab:adata}--\ref{tab:idata} in \ref{sec:data}.
Note that estimates for $\avresq/\avrgsq$ are included because the positive
correlation between the two observables results in variance reduction.

\section{Analysis}
\label{sec:analysis}

Finding a good method to analyze our data and obtain an accurate estimate for
$w^*$  is a difficult problem. The heart of the difficulty is that we have
conflicting goals that need to be accommodated, namely we wish to reduce the
influence of unfitted corrections to scaling and thus go to the large-$N$ limit,
and we wish to study the Domb-Joyce model in the vicinity of $w \approx w^*$,
but at the same time we want a large signal for the
leading-correction-to-scaling term, which necessitates collecting data for intermediate values of $N$, and
for values of $w$ which are quite far from $w^*$.

\begin{figure}[tb]
\begin{center}
\includegraphics[width=0.8\textwidth]{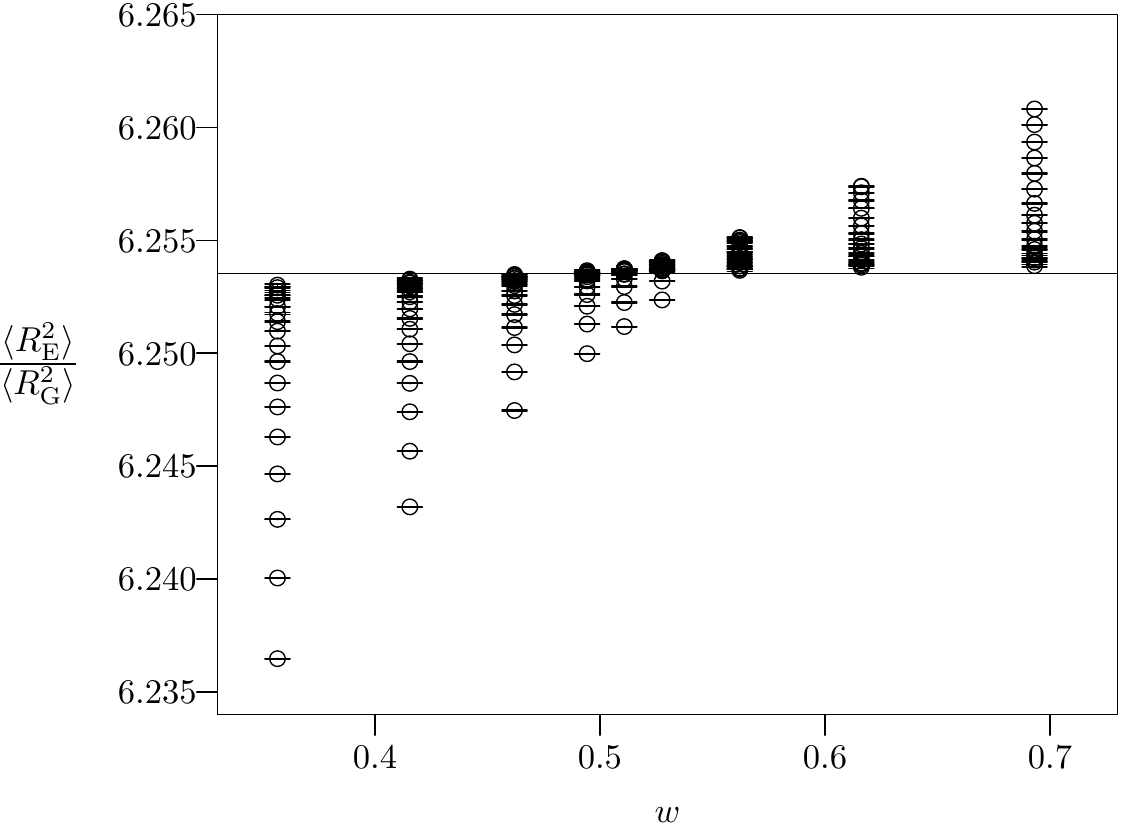}
\end{center}
\vspace{-4ex}
    \caption{$\avresq/\avrgsq$ as a function of $w$ for $1000 \leq N \leq 10^7$. The smaller values of $N$ have larger corrections to scaling, and so deviate to a greater extent from the limiting value. We show a horizontal line for the best estimate of $\dde/\ddg = 6.253531$.
\label{fig:re2rg2}}
\end{figure}

The most robust and accurate method we could devise involves the use of
information about the value of $\dde/\ddg = 6.253531(10)$ from simulations of
SAWs in the large-$N$ limit~\cite{Clisby2016HydrodynamicRadiusForSAWs}. For
fixed $N$, we found the value of $w(N)$ such that $\avresq_N/\avrgsq_N =
\dde/\ddg$ via a quartic fit.  These fits are perfect, in the sense that the
reduced $\chi^2$ value for the fits are approximately one indicating that the
model accurately describes the data, and so we are confident that the resulting
estimates for $w(N)$ are reliable.  We found that the error bars for $w(N)$
became too large to be useful for $N \geq 10^6$.  One can see the reason for
this in Fig.~\ref{fig:re2rg2}: as $N$ increases, $N^{-\Delta_1}$ decreases, the
amplitude of deviations from the limiting value of $\dde/\ddg$ become smaller,
and so errors on the estimates for $w(N)$ increase.

To gain an understanding of the convergence behavior of our estimates $w(N)$ we
examine the equation we are solving when the next correction-to-scaling term
with exponent $-1$ is included.
The expected asymptotic form for the ratio $\avresq_N/\avrgsq_N$ is
\begin{align}
    \frac{\avresq_N}{\avrgsq_N} &=
    \frac{\dde}{\ddg} \left(1 + \frac{f(w)}{N^{\Delta_1}} + \frac{g(w)}{N} + \cdots  \right).
    \label{eq:rerg}
\end{align}
By solving $\avresq_N/\avrgsq_N = \dde/\ddg$ at finite $N$, we find the value of $w$ which forces the correction-to-scaling
terms to be zero in aggregate, i.e. we are solving
\begin{align}
    \frac{f(w)}{N^{\Delta_1}} + \frac{g(w)}{N} &= 0,
\end{align}
neglecting higher order corrections. Now, $N$ is large, and we expect the
solution for $w$ to be close to $w^*$, and so we take $w = w^* + \Delta w$, and
drop subleading terms:
\begin{align}
    f(w^*+\Delta w) &= -g(w^*+\Delta w) N^{\Delta_1-1},
   \\ f(w^*) + \Delta w f^\prime(w^*) &= -g(w^*) N^{\Delta_1-1},
    \\ \Delta w  &= -\frac{g(w^*)}{f^\prime(w^*)} N^{\Delta_1-1}.
\end{align}
Thus we expect the deviation of our estimates $w(N)$ from the limiting value
$w^*$ to be $\Delta w = O(N^{\Delta_1-1})$.
One quite significant issue with this assumption is that there are competing
next-to-leading correction terms, namely the analytic $O(N^{-1})$ correction
term, the $O(N^{-2\Delta_1})$ correction term, and a further non-analytic
correction term $O(N^{-\Delta_2})$ for which it is believed that $\Delta_2
\approx 1$.
We will assume for now that the deviation is well described by 
$\Delta w = O(N^{\Delta_1-1})$, but will touch on this point again later in
Sec.~\ref{sec:conclusion}.

\begin{figure}[htb]
\begin{center}
\includegraphics[width=0.8\textwidth]{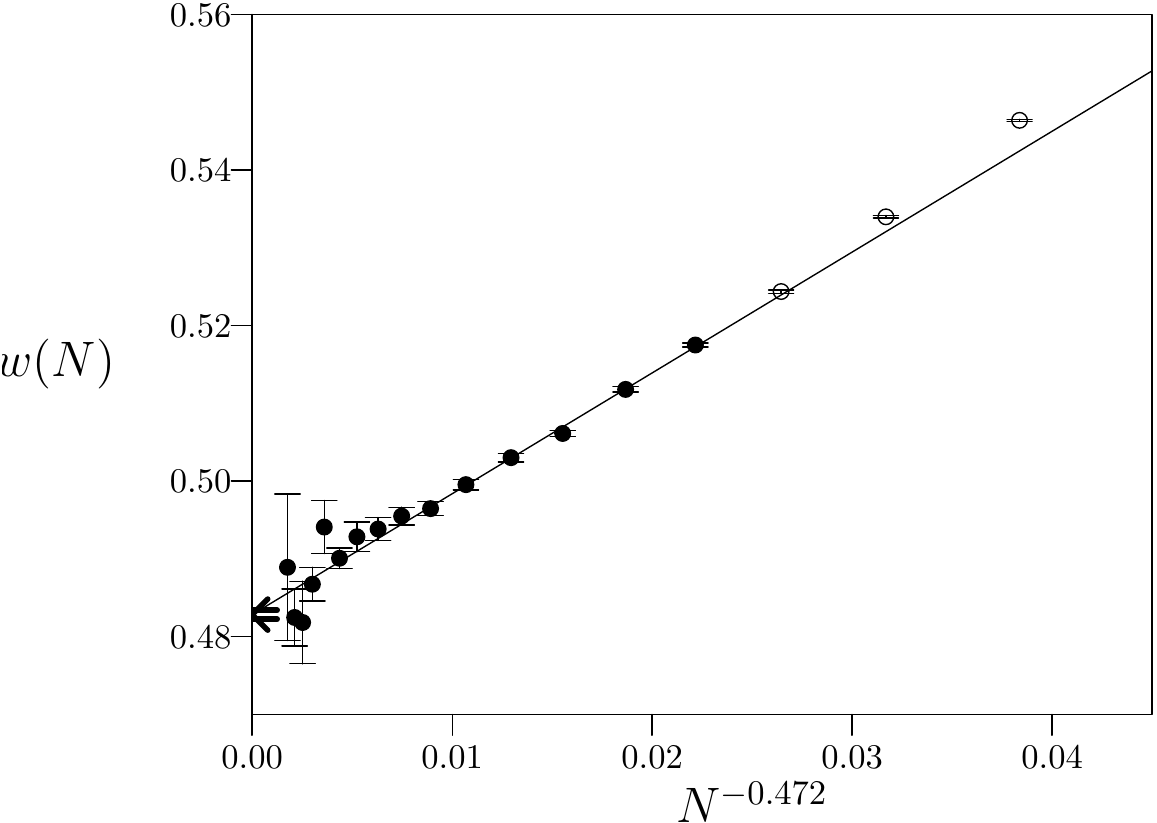}
\end{center}
\vspace{-4ex}
    \caption{Estimates for $w(N)$ for $1000 \leq N \leq 680000$ plotted against
    $N^{\Delta_1 - 1}$ with $\Delta_1 = 0.528$. The estimate for $w^* =
    0.48284(58)$, which is obtained by performing a linear fit of the solid data
    points, is shown on the vertical axis.
\label{fig:w}}
\end{figure}

Now that we have $w(N)$, we plot the estimates against $N^{\Delta_1-1}$ with
$\Delta_1 = 0.528$ in Fig.~\ref{fig:w}.  We performed linear fits of these data,
finding that the fit is excellent for
$N$ in the range $3200 \leq N \leq 680000$, with reduced $\chi^2$ approximately one. The resulting estimate of $w^* =
0.48284(58)$ is shown in the plot on the vertical axis.
Note that the estimates here for $w(N)$ are independent of each other, and hence
it makes sense to perform fits and then derive a statistical error estimate for
$w^*$, in contrast to, say, estimates of $\Delta_1$ in Fig.~10 of
\cite{Clisby2016HydrodynamicRadiusForSAWs} which form a correlated sequence for
which a linear fit no longer gives a meaningful statistical error.

Our estimate for $w^*$ is biased via the choice of values for $\dde/\ddg$
and $\Delta_1$. Although we have only shown estimates in
Fig.~\ref{fig:w} which are biased with the central values of estimates for
$\dde/\ddg$ and $\Delta_1$, we have varied these values within their confidence
intervals.
We find that varying $\dde/\ddg = 6.253531(10)$ within the confidence interval
$\pm 0.000010$ causes a variation in the estimate of $w^*$ of $\pm 0.0011$.
Varying $\Delta_1 = 0.528(8)$ within the confidence interval
$\pm 0.008$ causes a variation in the estimate of $w^*$ of $\pm 0.00034$.
Thus the largest source of error comes from our earlier estimate of 
$\dde/\ddg$.

We combine the statistical error of 0.00058 with the errors due to biasing as if
they were statistically independent sources of error, giving $\sigma(w^*) =
\sqrt{0.00058^2+0.0011^2+0.00034^2} = 0.0013$. Thus our final estimate,
incorporating all known sources of error, is 
$w^* = 0.4828(13)$.

Finally, we note that for the analysis we essentially have to take 2 limits, $N
\rightarrow \infty$, and $w \rightarrow w^*$. The previously described method
involves first solving for $w(N)$ (taking the limit $w \rightarrow w^*$ at
finite $N$), and then taking the limit $w^* = \lim_{N \rightarrow \infty} w(N)$.
We also tried taking the limit in the reverse order as follows.
We fitted data for $\avresq$ and $\avrgsq$ with biased values
for $\nu$ and $\Delta_1$, obtaining sequences of estimates for $\aae(w)$ and
$\aag(w)$ which were their limiting (large $N$) values. We then solved the
equations $\aae(w^*) = 0$ and $\aag(w^*)$ for $w^*$.  Unfortunately, 
we found that this process gave significantly larger error bars, and so we will
not report the details.

\section{Discussion and conclusion}
\label{sec:conclusion}

We have included all of the data here not only for completeness, but also with
the hope that an ingenious method of analysis may yet be found which can extract
more information from the data in Tables~\ref{tab:adata}--\ref{tab:idata}. We
note that a more accurate value for $\dde/\ddg$ would result in a
correspondingly more accurate estimate for $w^*$; it may be the case that such
an improved estimate comes from a simulation of the Domb-Joyce model with the
value for $w^*$ reported in this paper!

As mentioned in the previous section, the error bar for our estimate of $w^*$
relies on the fact that the next-to-leading correction term is well approximated
by a single $O(N^{-1})$ term, despite the fact that there may be competing
terms. The observation that the data in Fig.~\ref{fig:w} is well described by a
straight line gives us some confirmation that this is true, but nonetheless we
are mindful that the competing terms may result in a subtle error that could be
larger than our reported confidence interval for $w^*$. The only way we can
think of to mitigate this possible source of error is to obtain better data for
larger $N$, which would allow for easier extrapolation. But, this is a hard
problem: statistical errors are relatively larger for large $N$, so it is
unclear how much the situation would be improved by doing this. 

The use of the improved version of the Domb-Joyce model is undoubtedly an
attractive choice for many problems in polymer physics as it allows for faster
convergence. We note that there is another alternative to improve convergence
for particular observables by reducing the amplitude of the leading
correction-to-scaling term.  This has been done for the Ising
model~\cite{Hasenbusch20073dDilutedIsingImprovedObservable,Hasenbusch2010Finitesizescaling},
and more recently for self-avoiding
walks~\cite{Clisby2016HydrodynamicRadiusForSAWs,Clisby2017Scale-freeGammaSAWsArxiv}. The
manner in which the amplitude of the leading correction-to-scaling term is
reduced is independent between the two methods, and so in fact it is possible to
use both methods simultaneously and compound the effect. Thus, a study of the
Domb-Joyce model at $w = w^*$ for the improved observable $\rimp = \resq - 4.478
\rgsq$~\cite{Clisby2016HydrodynamicRadiusForSAWs} would be expected to have a
completely negligible leading correction-to-scaling term.

Finally, we give our best estimate for the value of $w$ which gives an improved
model as 
$w^* = 0.4828(13)$; we hope that it proves useful in future studies of the
venerable Domb-Joyce model.

\section*{Acknowledgements}
  Support from the Australian Research Council under
  the Future Fellowship scheme (project number FT130100972) and
    Discovery scheme (project
  number DP140101110) is gratefully acknowledged. 

\appendix

\section{Numerical data}
\label{sec:data}

\setlength{\belowcaptionskip}{0pt}

\begin{table}[H]
    \caption{Estimates of $\avresq$, $\avrgsq$, and $\avresq/\avrgsq$ for $e^{-w} = 0.50$.}
\label{tab:adata}
\begin{center}
\begin{tabular}{rllllll} 
\hline
    \multicolumn{1}{r}{$N$ \tstrut \bstrut}
    & \multicolumn{1}{c}{$\avresq$} & \multicolumn{1}{c}{$\avrgsq$}
    & \multicolumn{1}{c}{$\avresq/\avrgsq$} \\
\hline
1000 & 2.946985(15)\ee{3} & 4.707032(22)\ee{2} & 6.260814(16) \\
1500 & 4.749036(26)\ee{3} & 7.586170(38)\ee{2} & 6.260123(17) \\
2200 & 7.452665(44)\ee{3} & 1.1906453(66)\ee{3} & 6.259349(18) \\
3200 & 1.1581241(72)\ee{4} & 1.850439(11)\ee{3} & 6.258644(19) \\
4600 & 1.774760(12)\ee{4} & 2.836003(18)\ee{3} & 6.257964(21) \\
6800 & 2.810498(18)\ee{4} & 4.491575(28)\ee{3} & 6.257267(21) \\
10000 & 4.423265(31)\ee{4} & 7.069721(48)\ee{3} & 6.256633(22) \\
15000 & 7.125163(55)\ee{4} & 1.1389121(84)\ee{4} & 6.256113(24) \\
22000 & 1.1177829(89)\ee{5} & 1.786807(14)\ee{4} & 6.255755(25) \\
32000 & 1.736461(15)\ee{5} & 2.775938(24)\ee{4} & 6.255402(28) \\
46000 & 2.660357(25)\ee{5} & 4.253144(39)\ee{4} & 6.255037(29) \\
68000 & 4.211882(43)\ee{5} & 6.733903(66)\ee{4} & 6.254741(31) \\
100000 & 6.627622(31)\ee{5} & 1.0596410(49)\ee{5} & 6.254592(15) \\
150000 & 1.067418(14)\ee{6} & 1.706677(22)\ee{5} & 6.254363(39) \\
220000 & 1.6743015(91)\ee{6} & 2.677075(14)\ee{5} & 6.254219(17) \\
320000 & 2.600776(36)\ee{6} & 4.158496(57)\ee{5} & 6.254126(45) \\
460000 & 3.984107(25)\ee{6} & 6.370459(39)\ee{5} & 6.254035(19) \\
680000 & 6.30721(10)\ee{6} & 1.008529(17)\ee{6} & 6.253874(50) \\
1000000 & 9.92384(17)\ee{6} & 1.586832(27)\ee{6} & 6.253872(53) \\
3200000 & 3.89384(17)\ee{7} & 6.22638(27)\ee{6} & 6.25378(15) \\
10000000 & 1.485785(85)\ee{8} & 2.37579(13)\ee{7} & 6.25386(16) \\
\hline
\end{tabular}
\end{center}
\end{table}

\begin{table}[H]
    \caption{Estimates of $\avresq$, $\avrgsq$, and $\avresq/\avrgsq$ for $e^{-w} = 0.54$.}
\label{tab:bdata}
\begin{center}
\begin{tabular}{rllllll} 
\hline
    \multicolumn{1}{r}{$N$ \tstrut \bstrut}
    & \multicolumn{1}{c}{$\avresq$} & \multicolumn{1}{c}{$\avrgsq$}
    & \multicolumn{1}{c}{$\avresq/\avrgsq$} \\
\hline
1000 & 2.850082(14)\ee{3} & 4.554742(21)\ee{2} & 6.257394(16) \\
1500 & 4.591680(25)\ee{3} & 7.338035(37)\ee{2} & 6.257370(17) \\
2200 & 7.204270(43)\ee{3} & 1.1513746(66)\ee{3} & 6.257103(19) \\
3200 & 1.1193133(72)\ee{4} & 1.788966(11)\ee{3} & 6.256760(20) \\
4600 & 1.715068(12)\ee{4} & 2.741294(18)\ee{3} & 6.256416(21) \\
6800 & 2.715633(18)\ee{4} & 4.340859(27)\ee{3} & 6.255981(21) \\
10000 & 4.273541(31)\ee{4} & 6.831499(48)\ee{3} & 6.255641(22) \\
15000 & 6.883533(52)\ee{4} & 1.1004328(81)\ee{4} & 6.255296(23) \\
22000 & 1.0798013(90)\ee{5} & 1.726294(14)\ee{4} & 6.255027(26) \\
32000 & 1.677379(15)\ee{5} & 2.681733(23)\ee{4} & 6.254834(27) \\
46000 & 2.569767(25)\ee{5} & 4.108577(39)\ee{4} & 6.254640(31) \\
68000 & 4.068303(43)\ee{5} & 6.504673(68)\ee{4} & 6.254432(32) \\
100000 & 6.401507(31)\ee{5} & 1.0235392(49)\ee{5} & 6.254287(15) \\
150000 & 1.030977(13)\ee{6} & 1.648477(21)\ee{5} & 6.254119(40) \\
220000 & 1.6171028(89)\ee{6} & 2.585713(14)\ee{5} & 6.253992(17) \\
320000 & 2.511851(38)\ee{6} & 4.016416(59)\ee{5} & 6.253962(45) \\
460000 & 3.847937(24)\ee{6} & 6.152863(38)\ee{5} & 6.253897(19) \\
680000 & 6.091492(99)\ee{6} & 9.74046(16)\ee{5} & 6.253806(49) \\
1000000 & 9.58453(18)\ee{6} & 1.532626(28)\ee{6} & 6.253662(53) \\
3200000 & 3.76039(17)\ee{7} & 6.01306(26)\ee{6} & 6.25371(14) \\
10000000 & 1.434767(84)\ee{8} & 2.29433(12)\ee{7} & 6.25353(17) \\
\hline
\end{tabular}
\end{center}
\end{table}

\begin{table}[H]
    \caption{Estimates of $\avresq$, $\avrgsq$, and $\avresq/\avrgsq$ for $e^{-w} = 0.57$.}
\label{tab:cdata}
\begin{center}
\begin{tabular}{rllllll} 
\hline
    \multicolumn{1}{r}{$N$ \tstrut \bstrut}
    & \multicolumn{1}{c}{$\avresq$} & \multicolumn{1}{c}{$\avrgsq$}
    & \multicolumn{1}{c}{$\avresq/\avrgsq$} \\
\hline
1000 & 2.775558(13)\ee{3} & 4.437707(20)\ee{2} & 6.254487(15) \\
1500 & 4.470436(25)\ee{3} & 7.147028(38)\ee{2} & 6.254958(17) \\
2200 & 7.012679(41)\ee{3} & 1.1211100(63)\ee{3} & 6.255121(19) \\
3200 & 1.0894001(68)\ee{4} & 1.741618(10)\ee{3} & 6.255103(20) \\
4600 & 1.669035(11)\ee{4} & 2.668318(17)\ee{3} & 6.255007(21) \\
6800 & 2.642508(18)\ee{4} & 4.224700(28)\ee{3} & 6.254902(21) \\
10000 & 4.158105(30)\ee{4} & 6.647938(46)\ee{3} & 6.254730(23) \\
15000 & 6.697016(52)\ee{4} & 1.0707311(80)\ee{4} & 6.254620(24) \\
22000 & 1.0504808(91)\ee{5} & 1.679572(14)\ee{4} & 6.254455(26) \\
32000 & 1.631763(15)\ee{5} & 2.609020(23)\ee{4} & 6.254314(28) \\
46000 & 2.499768(24)\ee{5} & 3.996946(38)\ee{4} & 6.254194(30) \\
68000 & 3.957434(40)\ee{5} & 6.327736(64)\ee{4} & 6.254107(31) \\
100000 & 6.226822(31)\ee{5} & 9.956529(48)\ee{4} & 6.254010(15) \\
150000 & 1.002843(13)\ee{6} & 1.603557(21)\ee{5} & 6.253865(39) \\
220000 & 1.5729401(87)\ee{6} & 2.515147(14)\ee{5} & 6.253870(17) \\
320000 & 2.443194(35)\ee{6} & 3.906756(56)\ee{5} & 6.253766(45) \\
460000 & 3.742700(23)\ee{6} & 5.984739(37)\ee{5} & 6.253740(19) \\
680000 & 5.92494(10)\ee{6} & 9.47434(16)\ee{5} & 6.253670(51) \\
1000000 & 9.32244(17)\ee{6} & 1.490718(27)\ee{6} & 6.253659(54) \\
3200000 & 3.65761(17)\ee{7} & 5.84878(27)\ee{6} & 6.25362(15) \\
10000000 & 1.395505(75)\ee{8} & 2.23149(12)\ee{7} & 6.25368(16) \\
\hline
\end{tabular}
\end{center}
\end{table}

\begin{table}[H]
    \caption{Estimates of $\avresq$, $\avrgsq$, and $\avresq/\avrgsq$ for $e^{-w} = 0.59$.}
\label{tab:ddata}
\begin{center}
\begin{tabular}{rllllll} 
\hline
    \multicolumn{1}{r}{$N$  \tstrut \bstrut}
    & \multicolumn{1}{c}{$\avresq$} & \multicolumn{1}{c}{$\avrgsq$}
    & \multicolumn{1}{c}{$\avresq/\avrgsq$} \\
\hline
1000 & 2.724788(14)\ee{3} & 4.358016(20)\ee{2} & 6.252359(16) \\
1500 & 4.387885(24)\ee{3} & 7.017033(36)\ee{2} & 6.253191(17) \\
2200 & 6.882280(41)\ee{3} & 1.1005222(62)\ee{3} & 6.253649(19) \\
3200 & 1.0690098(69)\ee{4} & 1.709346(10)\ee{3} & 6.253911(20) \\
4600 & 1.637665(11)\ee{4} & 2.618552(18)\ee{3} & 6.254086(22) \\
6800 & 2.592544(20)\ee{4} & 4.145352(31)\ee{3} & 6.254098(23) \\
10000 & 4.079305(34)\ee{4} & 6.522653(52)\ee{3} & 6.254058(25) \\
15000 & 6.569770(57)\ee{4} & 1.0504836(88)\ee{4} & 6.254044(27) \\
22000 & 1.0304625(98)\ee{5} & 1.647691(15)\ee{4} & 6.253980(29) \\
32000 & 1.600626(16)\ee{5} & 2.559396(24)\ee{4} & 6.253921(31) \\
46000 & 2.451997(26)\ee{5} & 3.920766(40)\ee{4} & 6.253873(34) \\
68000 & 3.881737(46)\ee{5} & 6.206968(70)\ee{4} & 6.253838(36) \\
100000 & 6.107504(76)\ee{5} & 9.76620(12)\ee{4} & 6.253715(37) \\
150000 & 9.83587(13)\ee{5} & 1.572791(21)\ee{5} & 6.253766(40) \\
220000 & 1.542762(21)\ee{6} & 2.466964(33)\ee{5} & 6.253689(43) \\
320000 & 2.396316(36)\ee{6} & 3.831840(56)\ee{5} & 6.253695(45) \\
460000 & 3.670882(59)\ee{6} & 5.869984(93)\ee{5} & 6.253649(49) \\
680000 & 5.811180(98)\ee{6} & 9.29243(16)\ee{5} & 6.253672(50) \\
1000000 & 9.14307(16)\ee{6} & 1.462063(25)\ee{6} & 6.253540(54) \\
3200000 & 3.58716(20)\ee{7} & 5.73637(32)\ee{6} & 6.25336(15) \\
10000000 & 1.368788(75)\ee{8} & 2.18877(12)\ee{7} & 6.25369(18) \\
\hline
\end{tabular}
\end{center}
\end{table}

\begin{table}[H]
    \caption{Estimates of $\avresq$, $\avrgsq$, and $\avresq/\avrgsq$ for $e^{-w} = 0.60$.}
\label{tab:edata}
\begin{center}
\begin{tabular}{rllllll} 
\hline
    \multicolumn{1}{r}{$N$ \tstrut \bstrut}
    & \multicolumn{1}{c}{$\avresq$} & \multicolumn{1}{c}{$\avrgsq$}
    & \multicolumn{1}{c}{$\avresq/\avrgsq$} \\
\hline
1000 & 2.699053(14)\ee{3} & 4.317670(21)\ee{2} & 6.251180(16) \\
1500 & 4.346023(24)\ee{3} & 6.951142(36)\ee{2} & 6.252243(17) \\
2200 & 6.816145(41)\ee{3} & 1.0900684(62)\ee{3} & 6.252951(19) \\
3200 & 1.0586576(69)\ee{4} & 1.692961(11)\ee{3} & 6.253292(21) \\
4600 & 1.621716(11)\ee{4} & 2.593297(17)\ee{3} & 6.253492(22) \\
6800 & 2.567270(18)\ee{4} & 4.105223(27)\ee{3} & 6.253668(22) \\
10000 & 4.039346(29)\ee{4} & 6.459106(46)\ee{3} & 6.253724(23) \\
15000 & 6.505188(52)\ee{4} & 1.0402078(80)\ee{4} & 6.253739(25) \\
22000 & 1.0203202(85)\ee{5} & 1.631538(13)\ee{4} & 6.253733(26) \\
32000 & 1.584838(14)\ee{5} & 2.534228(22)\ee{4} & 6.253733(28) \\
46000 & 2.427777(23)\ee{5} & 3.882134(37)\ee{4} & 6.253717(30) \\
68000 & 3.843233(40)\ee{5} & 6.145599(64)\ee{4} & 6.253634(33) \\
100000 & 6.047151(30)\ee{5} & 9.669732(47)\ee{4} & 6.253691(15) \\
150000 & 9.73841(13)\ee{5} & 1.557242(20)\ee{5} & 6.253630(42) \\
220000 & 1.5274664(86)\ee{6} & 2.442523(14)\ee{5} & 6.253642(17) \\
320000 & 2.372506(36)\ee{6} & 3.793808(58)\ee{5} & 6.253628(47) \\
460000 & 3.634416(23)\ee{6} & 5.811689(36)\ee{5} & 6.253631(19) \\
680000 & 5.753378(99)\ee{6} & 9.20023(16)\ee{5} & 6.253516(52) \\
1000000 & 9.05234(16)\ee{6} & 1.447554(26)\ee{6} & 6.253541(56) \\
3200000 & 3.55175(17)\ee{7} & 5.67980(26)\ee{6} & 6.25331(16) \\
10000000 & 1.355088(79)\ee{8} & 2.16690(13)\ee{7} & 6.25358(15) \\
\hline
\end{tabular}
\end{center}
\end{table}

\begin{table}[H]
    \caption{Estimates of $\avresq$, $\avrgsq$, and $\avresq/\avrgsq$ for $e^{-w} = 0.61$.}
\label{tab:fdata}
\begin{center}
\begin{tabular}{rllllll} 
\hline
    \multicolumn{1}{r}{$N$ \tstrut \bstrut}
    & \multicolumn{1}{c}{$\avresq$} & \multicolumn{1}{c}{$\avrgsq$}
    & \multicolumn{1}{c}{$\avresq/\avrgsq$} \\
\hline
1000 & 2.673077(14)\ee{3} & 4.276938(21)\ee{2} & 6.249979(16) \\
1500 & 4.303817(24)\ee{3} & 6.884686(36)\ee{2} & 6.251290(17) \\
2200 & 6.749264(41)\ee{3} & 1.0795231(62)\ee{3} & 6.252079(19) \\
3200 & 1.0482164(68)\ee{4} & 1.676450(10)\ee{3} & 6.252597(20) \\
4600 & 1.605641(11)\ee{4} & 2.567824(17)\ee{3} & 6.252927(22) \\
6800 & 2.541686(20)\ee{4} & 4.064615(30)\ee{3} & 6.253203(24) \\
10000 & 3.998813(33)\ee{4} & 6.394748(51)\ee{3} & 6.253277(25) \\
15000 & 6.439809(58)\ee{4} & 1.0298109(88)\ee{4} & 6.253390(28) \\
22000 & 1.0100599(97)\ee{5} & 1.615200(15)\ee{4} & 6.253466(30) \\
32000 & 1.568852(16)\ee{5} & 2.508752(25)\ee{4} & 6.253515(32) \\
46000 & 2.403254(26)\ee{5} & 3.843032(40)\ee{4} & 6.253535(33) \\
68000 & 3.804503(45)\ee{5} & 6.083705(69)\ee{4} & 6.253595(36) \\
100000 & 5.985813(74)\ee{5} & 9.57190(12)\ee{4} & 6.253530(37) \\
150000 & 9.63989(12)\ee{5} & 1.541516(19)\ee{5} & 6.253515(41) \\
220000 & 1.511978(22)\ee{6} & 2.417815(34)\ee{5} & 6.253489(44) \\
320000 & 2.348474(36)\ee{6} & 3.755363(56)\ee{5} & 6.253654(46) \\
460000 & 3.597410(57)\ee{6} & 5.752654(92)\ee{5} & 6.253478(49) \\
680000 & 5.694988(95)\ee{6} & 9.10673(15)\ee{5} & 6.253601(51) \\
1000000 & 8.96033(16)\ee{6} & 1.432841(25)\ee{6} & 6.253539(55) \\
3200000 & 3.51543(17)\ee{7} & 5.62146(28)\ee{6} & 6.25358(15) \\
10000000 & 1.341254(82)\ee{8} & 2.14478(13)\ee{7} & 6.25357(17) \\
\hline
\end{tabular}
\end{center}
\end{table}

\begin{table}[H]
    \caption{Estimates of $\avresq$, $\avrgsq$, and $\avresq/\avrgsq$ for $e^{-w} = 0.63$.}
\label{tab:gdata}
\begin{center}
\begin{tabular}{rllllll} 
\hline
    \multicolumn{1}{r}{$N$ \tstrut \bstrut}
    & \multicolumn{1}{c}{$\avresq$} & \multicolumn{1}{c}{$\avrgsq$}
    & \multicolumn{1}{c}{$\avresq/\avrgsq$} \\
\hline
1000 & 2.620372(13)\ee{3} & 4.194298(20)\ee{2} & 6.247461(16) \\
1500 & 4.217911(23)\ee{3} & 6.749552(36)\ee{2} & 6.249172(17) \\
2200 & 6.613457(40)\ee{3} & 1.0580911(60)\ee{3} & 6.250366(19) \\
3200 & 1.0269831(67)\ee{4} & 1.642874(10)\ee{3} & 6.251136(20) \\
4600 & 1.572935(11)\ee{4} & 2.516009(17)\ee{3} & 6.251709(22) \\
6800 & 2.489640(17)\ee{4} & 3.982046(27)\ee{3} & 6.252162(22) \\
10000 & 3.916771(29)\ee{4} & 6.264273(45)\ee{3} & 6.252556(23) \\
15000 & 6.307029(50)\ee{4} & 1.0086802(77)\ee{4} & 6.252754(24) \\
22000 & 9.891915(84)\ee{4} & 1.581950(13)\ee{4} & 6.252988(27) \\
32000 & 1.536343(14)\ee{5} & 2.456945(22)\ee{4} & 6.253063(29) \\
46000 & 2.353399(23)\ee{5} & 3.763530(36)\ee{4} & 6.253171(31) \\
68000 & 3.725361(39)\ee{5} & 5.957515(62)\ee{4} & 6.253214(32) \\
100000 & 5.861333(29)\ee{5} & 9.373172(46)\ee{4} & 6.253308(15) \\
150000 & 9.43908(13)\ee{5} & 1.509455(20)\ee{5} & 6.253308(40) \\
220000 & 1.4804725(85)\ee{6} & 2.367472(13)\ee{5} & 6.253389(17) \\
320000 & 2.299456(36)\ee{6} & 3.677109(57)\ee{5} & 6.253436(47) \\
460000 & 3.522448(22)\ee{6} & 5.632776(35)\ee{5} & 6.253486(19) \\
680000 & 5.576181(90)\ee{6} & 8.91697(14)\ee{5} & 6.253448(51) \\
1000000 & 8.77324(16)\ee{6} & 1.402950(26)\ee{6} & 6.253424(55) \\
3200000 & 3.44241(16)\ee{7} & 5.50497(25)\ee{6} & 6.25327(13) \\
10000000 & 1.313124(74)\ee{8} & 2.09986(12)\ee{7} & 6.25340(19) \\
\hline
\end{tabular}
\end{center}
\end{table}

\begin{table}[H]
    \caption{Estimates of $\avresq$, $\avrgsq$, and $\avresq/\avrgsq$ for $e^{-w} = 0.66$.}
\label{tab:hdata}
\begin{center}
\begin{tabular}{rllllll} 
\hline
    \multicolumn{1}{r}{$N$ \tstrut \bstrut}
    & \multicolumn{1}{c}{$\avresq$} & \multicolumn{1}{c}{$\avrgsq$}
    & \multicolumn{1}{c}{$\avresq/\avrgsq$} \\
\hline
1000 & 2.539064(13)\ee{3} & 4.066931(20)\ee{2} & 6.243193(16) \\
1500 & 4.085488(24)\ee{3} & 6.541324(36)\ee{2} & 6.245660(18) \\
2200 & 6.403834(40)\ee{3} & 1.0250392(61)\ee{3} & 6.247404(20) \\
3200 & 9.941852(65)\ee{3} & 1.591036(10)\ee{3} & 6.248665(21) \\
4600 & 1.522358(11)\ee{4} & 2.435917(17)\ee{3} & 6.249628(23) \\
6800 & 2.409223(17)\ee{4} & 3.854499(26)\ee{3} & 6.250417(22) \\
10000 & 3.789693(28)\ee{4} & 6.062478(44)\ee{3} & 6.251062(24) \\
15000 & 6.101844(51)\ee{4} & 9.760545(78)\ee{3} & 6.251540(25) \\
22000 & 9.568984(84)\ee{4} & 1.530557(13)\ee{4} & 6.251963(27) \\
32000 & 1.486076(14)\ee{5} & 2.376861(21)\ee{4} & 6.252264(29) \\
46000 & 2.276301(24)\ee{5} & 3.640626(36)\ee{4} & 6.252500(31) \\
68000 & 3.603156(39)\ee{5} & 5.762544(61)\ee{4} & 6.252718(33) \\
100000 & 5.668681(28)\ee{5} & 9.065749(45)\ee{4} & 6.252855(15) \\
150000 & 9.12854(13)\ee{5} & 1.459871(20)\ee{5} & 6.252979(42) \\
220000 & 1.4317105(82)\ee{6} & 2.289599(13)\ee{5} & 6.253105(17) \\
320000 & 2.223729(34)\ee{6} & 3.556177(56)\ee{5} & 6.253145(46) \\
460000 & 3.406320(22)\ee{6} & 5.447284(34)\ee{5} & 6.253246(20) \\
680000 & 5.392277(93)\ee{6} & 8.62312(15)\ee{5} & 6.253280(51) \\
1000000 & 8.48419(15)\ee{6} & 1.356741(24)\ee{6} & 6.253357(55) \\
3200000 & 3.32818(16)\ee{7} & 5.32210(25)\ee{6} & 6.25351(15) \\
10000000 & 1.269788(75)\ee{8} & 2.03055(11)\ee{7} & 6.25342(18) \\
\hline
\end{tabular}
\end{center}
\end{table}

\begin{table}[H]
    \caption{Estimates of $\avresq$, $\avrgsq$, and $\avresq/\avrgsq$ for $e^{-w} = 0.70$.}
\label{tab:idata}
\begin{center}
\begin{tabular}{rllllll} 
\hline
    \multicolumn{1}{r}{$N$ \tstrut \bstrut}
    & \multicolumn{1}{c}{$\avresq$} & \multicolumn{1}{c}{$\avrgsq$}
    & \multicolumn{1}{c}{$\avresq/\avrgsq$} \\
\hline
1000 & 2.425877(13)\ee{3} & 3.889822(20)\ee{2} & 6.236472(17) \\
1500 & 3.900841(23)\ee{3} & 6.251304(35)\ee{2} & 6.240044(19) \\
2200 & 6.111309(39)\ee{3} & 9.789621(60)\ee{2} & 6.242641(19) \\
3200 & 9.483658(64)\ee{3} & 1.5186841(97)\ee{3} & 6.244654(21) \\
4600 & 1.451779(11)\ee{4} & 2.324226(16)\ee{3} & 6.246290(23) \\
6800 & 2.296771(17)\ee{4} & 3.676234(25)\ee{3} & 6.247619(22) \\
10000 & 3.611923(28)\ee{4} & 5.780298(43)\ee{3} & 6.248680(23) \\
15000 & 5.814414(49)\ee{4} & 9.303614(76)\ee{3} & 6.249630(25) \\
22000 & 9.116720(82)\ee{4} & 1.458600(13)\ee{4} & 6.250322(27) \\
32000 & 1.415705(13)\ee{5} & 2.264775(21)\ee{4} & 6.250976(29) \\
46000 & 2.168223(22)\ee{5} & 3.468382(35)\ee{4} & 6.251396(31) \\
68000 & 3.431820(37)\ee{5} & 5.489370(58)\ee{4} & 6.251756(34) \\
100000 & 5.398749(28)\ee{5} & 8.635145(44)\ee{4} & 6.252065(16) \\
150000 & 8.69335(12)\ee{5} & 1.390401(18)\ee{5} & 6.252402(43) \\
220000 & 1.3633795(80)\ee{6} & 2.180512(13)\ee{5} & 6.252567(18) \\
320000 & 2.117499(34)\ee{6} & 3.386514(53)\ee{5} & 6.252739(48) \\
460000 & 3.243508(22)\ee{6} & 5.187204(34)\ee{5} & 6.252903(20) \\
680000 & 5.134404(91)\ee{6} & 8.21109(14)\ee{5} & 6.253013(52) \\
1000000 & 8.07806(16)\ee{6} & 1.291849(24)\ee{6} & 6.253099(56) \\
3200000 & 3.16904(18)\ee{7} & 5.06766(27)\ee{6} & 6.25346(15) \\
10000000 & 1.209063(77)\ee{8} & 1.93345(12)\ee{7} & 6.25340(19) \\
\hline
\end{tabular}
\end{center}
\end{table}

\normalsize



\providecommand{\newblock}{}

\end{document}